 \documentclass[preprint,review,12pt]{elsarticle}
\usepackage{graphicx}

\usepackage{amssymb}
\usepackage{amsmath}
\usepackage{bm}

\newcommand{\dr}{\partial_r}
\newcommand{\ket}[1]{\left|#1\right\rangle}
\newcommand{\bra}[1]{\left\langle#1\right|}
\newcommand{\svect}[2]{\left(\begin{smallmatrix}#1\\#2\end{smallmatrix}\right)}
\newcommand{\vect}[2]{\left(\begin{matrix}#1\\#2\end{matrix}\right)}
\newcommand{\mat}[4]{\left(\begin{smallmatrix}#1 & #2\\#3 & #4\end{smallmatrix}\right)}
\newcommand{\matl}[4]{\left(\begin{matrix}#1 & #2\\#3 & #4\end{matrix}\right)}

\journal{Nuclear Physics B}

\begin{document}

\begin{frontmatter}

\title{Zero-energy states bound to a magnetic $\pi$-flux vortex in a two-dimensional topological insulator}

 \author[label1,label2]{Andrej Mesaros}
\author[label2]{Robert-Jan Slager}
\author[label2]{Jan Zaanen}
\author[label2]{Vladimir Juri\v ci\' c}

\address[label1]{Department of Physics, Boston College, Chestnut Hill, Massachusetts 02467, USA}
\address[label2]{Instituut-Lorentz for Theoretical Physics, Universiteit Leiden, P.O. Box 9506, 2300 RA Leiden, The Netherlands}

\begin{abstract}
  We show that the existence of a pair of zero-energy modes bound to  a vortex carrying a $\pi$-flux is a generic feature of the topologically non-trivial phase of the $M-B$ model, which was introduced to describe the topological band insulator in HgTe quantum wells.
We explicitly find the form of the zero-energy states of the corresponding Dirac equation, which contains a novel momentum-dependent mass term and describes a generic topological transition in a band insulator.
  The obtained modes are exponentially localized in the vortex-core, with the dependence of characteristic length on the parameters of the model matching the dependence extracted from a lattice version of the model.
  We consider in full generality the short-distance regularization of the vector potential of the vortex, and show that a particular choice
  yields the modes localized and simultaneously regular at the origin. Finally, we also discuss a realization of two-dimensional spin-charge separation through the vortex zero-modes.
\end{abstract}


\end{frontmatter}


\section{Introduction}
\label{Intro}
Topological band insulators (TBIs) have recently opened a new frontier in both theoretical and experimental condensed-matter physics due to their peculiar
properties (extensive reviews are in Refs. \cite{TIreview,hasan-review}). These stem from the fundamental fact that TBIs are described by topological field theories \cite{TFTTI}, thereby taking the interest in studying them far outside the standard condensed-matter domain. The novelty of TBIs is their protection by time-reversal symmetry (TRS), which leads to a non-trivial topological ${\mathbb Z}_2$ invariant \cite{fu-kane} of their free electron crystalline band structure, and to a description in terms of the topological $BF$ theory in (2+1)D and (3+1)D~\cite{BF}. Striking consequences of their topological nature are the remarkable effective field theories describing the responses of TBIs, for instance, axion electrodynamics in (3+1)D \cite{response} and gravitational Chern-Simons in thermal response \cite{response,gravanomaly}. Such theories are governed by anomalies, and the possibility of their direct study, e.g. through the Witten effect of axion electrodynamics \cite{Witteneffect}, is of great interest and potential for both the high-energy and condensed matter communities. Further beyond the non-interacting case, there has been a proposal of (3+1)D fractional TBIs \cite{fracTI}, having TRS and an axion angle different from $0$ or $\pi$; these have a description in the form of deconfined non-Abelian gauge fields explicitly realized using holography \cite{holographyTI}.

The salient feature of TBIs, crucial for their characterization and detection, is that they are fully gapped in the bulk while possessing on their boundary gapless propagating modes protected by TRS. \cite{TIreview,hasan-review} The presence of TRS however limits, fundamentally and especially experimentally, the availability of robust probes of such bulk-boundary correspondence, aforementioned anomalous responses, and ${\mathbb Z}_2$ topological order itself. For instance, in (2+1)D TBIs, on which we are focusing from now on, the charge Hall response vanishes, and instead a much more involved TRS invariant quantum spin Hall (QSH) effect characterizes the topological phase. It has been understood through numeric studies that a $\pi$-flux vortex, which actually preserves TRS, can play exactly the role of a ${\mathbb Z}_2$ probe in a QSH insulator through appearance of topologically protected zero-modes. \cite{BF,TopProbes,dung-hai1,qi-zhang}

In this paper, we analytically study the properties of $\pi$-flux vortex in presence and absence of ${\mathbb Z}_2$ order. The general way by which we achieve this is using the $M-B$ model initially constructed to describe the HgTe quantum well QSH insulator. \cite{bernevig,konig} The salient and universal feature of the low-energy (continuum) version of this model is that it describes a topological phase transition between a trivial and non-trivial ${\mathbb Z}_2$ topological phase, through a massive Dirac-Schr\" odinger theory. This field theory, especially in the presence of a $U(1)$ vortex, has not been widely studied for its own sake. A peculiar property of this theory is that the presence of both linear and quadratic kinetic terms together with the ordinary Dirac mass term allows for a gap-closing transition which changes the Chern number of the bands and the ${\mathbb Z}_2$ invariant. The same theory turns out to harbor analytically solvable zero-modes tied to $\pi$-flux vortex, but only in the non-trivial phase.

The relationship of the $\pi$-flux modes to the QSH phase and the question of their protection are general problems in the context of zero-energy fermionic modes bound to a topological defect. Namely, as Aharonov and Casher have shown in Ref.\ \cite{aharonov-casher}, when non-relativistic (Schr\" odinger) fermions are coupled to a magnetic flux carrying $n$ flux quanta there are precisely $n$ zero-energy modes in the spectrum of the Hamiltonian. Later, Jackiw in Ref.\ \cite{jackiw-prd84} has demonstrated that a magnetic flux with $n$ flux quanta hosts precisely $n$ zero-energy modes when coupled to relativistic Dirac fermions, and they are related to an index theorem for the Dirac Hamiltonian defined on a compact space.\cite{ansourian}
On the other hand,  the existence of fermionic zero-modes bound to a vortex in the complex scalar order parameter has actually been established in one and two spatial dimensions in the pioneering works by Jackiw and Rebbi\cite{jackiw-rebbi} and Jackiw and Rossi\cite{jackiw-rossi}, respectively. Their existence is, at the deep mathematical level, tied to an index theorem that relates the spectral asymmetry of the corresponding Hamiltonian defined on an open space  and a topological invariant of the background scalar fields.\cite{weinberg}  At the same time, the results of Jackiw and Rebbi have been applied to the polyacetilene system through the Su-Schrieffer-Heeger model.\cite{SSH}
However, the connection between Jackiw-Rossi and the $M-B$ model, relevant for the quantum spin Hall system, has been shown only very recently in Ref.\ \cite{chamon}. Namely, these authors have demonstrated that the Jackiw-Rossi  model in the limit when Zeeman coupling and the chemical potential are large reduces to the $M-B$ model, and therefore the results presented in this work are also relevant for this problem. Moreover, a Hamiltonian of the same form as the $M-B$ Hamiltonian describes non-relativistic $p_x+ip_y$ superconductor, and the results of our work are thus relevant for this system as well. In particular, our solution for the zero-energy mode bound to a $\pi$-flux vortex in the quantum spin Hall state corresponds to the Majorana state in the core of a $\pi$-flux vortex in this topological superconductor. 

The flux-carrying vortex is by its definition singular in real space. It is well known that a Hamiltonian with singular potential (e.g. Aharonov-Bohm flux vortex~\cite{Persson:2006p260,melikyan}, Coulomb potential~\cite{Fulop:2007p3442,KhalilovSAE,ParkSAE}, delta function potential~\cite{Jackiw:1991p1161}), once  made Hermitian through a self-adjoint extension \cite{Weidmann}, can exhibit finite or zero-energy bound states, even if the original Hamiltonian was scale-free. Therefore, the question of regularizing the vortex singularity, and thereby completely defining a Hermitian fermionic theory, becomes physically relevant. It fixes the real space profile of zero-modes at the vortex and the scattering phase shift there. This problem has been considered for (2+1)D ~\cite{Persson:2006p260,2dDiracSAE}, and (3+1)D~\cite{3dDiracSAE} Dirac particles, as well as quasiparticles in superconductors~\cite{melikyan}. The Dirac-Schr\" odinger theory of the $M-B$ model has not been studied in this context before, and we find here the general form of the regularized theory in presence of a magnetic $\pi$ flux vortex. Furthermore, we show that a regularization corresponding to a thin solenoid limit surprisingly leads to localized zero-modes that vanish at the origin.

In this paper, we first explicitly introduce the tight-binding and continuum $M-B$ models. Then we use the continuum $M-B$ model to analytically show that the $\pi$-flux vortex hosts precisely a pair of exponentially localized zero-energy modes, and therefore the states found numerically in Refs.\ \cite{dung-hai2, qi-zhang} are indeed a generic feature of the $M-B$ model. Moreover, the relationship of these midgap states with the topological properties of the quantum spin Hall state are also considered. Namely, we show that these modes, in fact, exist in the entire range of parameters describing topologically non-trivial phase in the $M-B$ model with the gap opening near the zero momentum ($\Gamma$ point) in the Brillouin zone. For obtaining the explicit form of the midgap states, a short-distance regularization of the Hamiltonian is necessary due to the singularity of the vortex potential. We show that a particular choice, corresponding to a limit of infinitely thin solenoid, yields the modes non-singular at the origin. We then consider in full generality the problem of consistent regularization of the vortex potential, and show that the corresponding self-adjoint extensions are given in terms of parameters described by $U(2)$ matrices. Next, we test the prediction for the localization length of the zero-modes obtained from the continuum theory as a function of the transition driving parameter by comparing it to the results from a lattice regularization of the model. Finally, we discuss the quantum numbers of the obtained zero-energy states.


\section{The $M-B$ tight-binding model on the square lattice}


We begin by considering a minimal tight-binding model proposed to describe a two-band quantum spin Hall insulator \cite{bernevig}
\begin{equation}\label{eq:tight-binding}
{\mathcal H}=\sum_{{\bf k}}\Psi^\dagger({\bf k})\left(\begin{array}{cc}H({\bf k})& 0\\
0 & H^*(-{\bf k})\end{array}\right)\Psi({\bf k})
\end{equation}
where $\Psi^\top=(u_\uparrow,v_\uparrow,u_\downarrow,v_\downarrow)\equiv(\Psi_\uparrow,\Psi_\downarrow)$, with $u$ and $v$ representing two low-energy orbitals. 
The upper and the lower blocks in the Hamiltonian are related by time-reversal symmetry, and $H({\bf k})$ acting in the orbital space has the form
\begin{equation}\label{eq:upper-ham}
H({\bf k})=\sigma_\mu d_\mu({\bf k}),
\end{equation}
where
$\sigma_\mu$, $\mu=1,2,3$, are the Pauli matrices, $d_{1,2}({\bf k})=A \sin k_{x,y}$, and $d_3=M-2B(2-\cos k_x-\cos k_y)$, the length unit is set by lattice constant $a=1$, and summation over repeated indices is assumed hereafter.
We also set $\hbar=c=e=1$ in the following, unless otherwise stated. Since the above Hamiltonian has spectrum $E({\bf k})=\sqrt{d_\mu d_\mu}$ doubly degenerate in spin space, the band gap closes at the $\Gamma$-point (${\bf k}=0$) in the Brillouin zone when the value of model parameters is $M/B=0$. We will only consider the range of parameters $0<M/B<4$ in what follows, so that the spectrum is gapped. This lattice model then describes a topologically nontrivial state with a Kramers' pair of counterpropagating modes on the edge of the system leading to a quantized spin Hall conductance\cite{konig} $\sigma_{xy}^{S}=2\frac{e^2}{h}$. For negative values of $M/B$, the model describes a trivial insulator ($\sigma_{xy}^S=0$).

\section{The Dirac-Schr\" odinger continuum theory}

By taking the continuum, i.e. large wavelength limit ($|{\bf k}|\ll 1$) of \eqref{eq:upper-ham}, we arrive at a Dirac Hamiltonian which besides the ordinary Dirac mass term ($M$) contains a Schr\" odinger kinetic term ($B$)
\begin{equation}\label{eq:cont-ham}
H_{\rm eff}({\bf k})=i\gamma_0\gamma_i k_i+(M-B{\bf k}^2)\gamma_0,
\end{equation}
where the four-dimensional $\gamma$-matrices are given by $\gamma_0=\sigma_3\otimes\tau_0$, $\gamma_1=\sigma_2\otimes\tau_3$, and $\gamma_2=-\sigma_1\otimes\tau_0$. Here, Pauli matrices $\{\tau_0,\tau_\mu\}$ act in spin space, with $\sigma_0,\tau_0$ as the $2\times 2$ identity matrices. The $\gamma$-matrices satisfy canonical anticommutation relations $\{\gamma_\alpha,\gamma_\beta\}=2\delta_{\alpha\beta}$, with $\alpha,\beta=0,1,2$.
 Notice that it is enough to focus on a single spin projection, since the two spin projections are related by the time-reversal operator $T=-i\tau_2K$, with $K$ as the complex conjugation. For convenience, we have divided through Eq.~\eqref{eq:tight-binding} by the energy scale $A$ (the lattice intersite hopping energy) and by the length-scale $a$, before redefining $M/(A a)\rightarrow M$, $B a/A\rightarrow B$, so that the continuum theory Eq.~\eqref{eq:cont-ham} has a dimensionless parameter $MB$ and parameters $B$ and $\sqrt{B/M}$ with dimension of length (we revert to lattice units for comparison to the tight binding model in Section \ref{sec:tb}).


\section{Zero-energy states}

Let us now consider the effect of the magnetic $\pi$-flux inserted into the system. As usually, the vector potential is coupled to the electronic degrees of freedom through the minimal substitution, ${\bf k}\rightarrow{\bf k}+{\bf A}$, and the Hamiltonian (\ref{eq:cont-ham}) for spin up electrons assumes the form
\begin{equation}\label{eq:ham-A}
H_{\rm eff}({\bf k},{\bf A})=\sigma_i (k_i+A_i)+[M-B({\bf k}+{\bf A})^2]\sigma_3.
\end{equation}
The vector potential
\begin{equation}\label{eq:A}
{\bf A}=\frac{-y{\bf e}_x+x{\bf e}_y}{2r^2}
\end{equation}
represents the magnetic vortex carrying the flux $\Phi=\pi$. Notice that the spin down electrons are coupled to the $\pi$-flux with the opposite sign because of the time-reversal symmetry, and in that
respect $\pi$-flux thus acts on the spin components as a pseudomagnetic vortex in graphene does on the two valley degrees of freedom.\cite{graphBerry,graphcones,fullerene} Of course, the time-reversal invariance of the Hamiltonian (\ref{eq:ham-A}) is present only when the flux corresponding to the vector potential ${\bf A}$ is equal to $\pi$ or $0$.

We now show that the above Hamiltonian possesses precisely one bulk zero-energy state with spin up.
Time-reversal symmetry then implies the
existence of the zero-energy state for electrons with spin down.
Expressing the Hamiltonian (\ref{eq:ham-A}) in polar coordinates $(r,\varphi)$, taking into account that ${\bf A}=(1/2r){\bf e}_\varphi$, we obtain
\begin{eqnarray}\label{eq:ham-A-explicit}
H_{\rm eff}&=&-i{ e}^{-i\varphi}\left[\partial_r-\frac{i}{r}{\tilde\partial}_\varphi\right]\sigma_ {+}-i{ e}^{i\varphi}\left[\partial_r+\frac{i}{r}{\tilde\partial}_\varphi\right]\sigma_{-}\nonumber\\
&+&\left[M+B\left(\partial_r^2+\frac{1}{r}\partial_r+\frac{1}{r^2}{\tilde\partial}_\varphi^2\right)\right]\sigma_3,
\end{eqnarray}
where ${\tilde\partial}_\varphi\equiv\partial_\varphi+i/2$, and $\sigma_\pm\equiv(\sigma_1\pm i\sigma_2)/2$. It is easy to see that in case of an arbitrary flux 
$\Phi$, the Hamiltonian (\ref{eq:ham-A}) also acquires the form  (\ref{eq:ham-A-explicit}), but with the operator ${\tilde\partial}_\varphi=\partial_\varphi+i\Phi/2\pi$.

In the presence of a vortex carrying a $\pi$-flux, we seek the zero-energy modes of the form
\begin{equation}\label{eq:zero-energy-spinor}
\Psi(r,\varphi)=\left(\begin{array}{cc}e^{i(l-1)\varphi}u_{l-1}(r)\\
e^{il\varphi}v_l(r)\end{array} \right),
\end{equation}
where $l\in \mathbb{Z}$ is the angular momentum quantum number, and the functions $u,v$ are the solutions of the following equations
\begin{eqnarray}
&&\Delta_{l-\frac{1}{2}}u_{l-1}(r)-i\left(\partial_r+\frac{l+\frac{1}{2}}{r}\right)v_l(r)=0\label{eq:zero-1}\\
&&i\left(\partial_r-\frac{l-\frac{1}{2}}{r}\right)u_{l-1}(r)+\Delta_{l+\frac{1}{2}}v_l(r)=0\label{eq:zero-2}.
\end{eqnarray}
Here the operator $\Delta_l$ is defined as
\begin{equation}\label{eq:Delta-def}
\Delta_l\equiv M+B\left(\partial_r^2+\frac{1}{r}\partial_r-\frac{l^2}{r^2}\right)\equiv M+B{\cal O}_l.
\end{equation}
Acting on Eq.\ (\ref{eq:zero-1}) with the operator $\Delta_{l+\frac{1}{2}}$, and using the identity
\begin{equation}
[\Delta_l,\partial_r+\frac{l}{r}]=-(2l-1)\frac{B}{r^2}\left(\partial_r+\frac{l}{r}\right),
\end{equation}
we can eliminate the function $v_l(r)$ from the same equation to obtain
\begin{equation}
\left(\Delta_{l+\frac{1}{2}}\Delta_{l-\frac{1}{2}}-{\cal O}_{l-\frac{1}{2}}+\frac{2Bl}{r^2}\Delta_{l-\frac{1}{2}}\right)u_{l-1}(r)=0.
\end{equation}
After some algebra, the above equation may be rewritten as
\begin{equation}\label{eq:zero-mode-u}
\left[M^2+(2MB-1){\cal O}_{l-\frac{1}{2}}+B^2{\cal O}_{l-\frac{1}{2}}^2\right]u_{l-1}(r)=0.
\end{equation}
This result may also be obtained by noting that if the spinor in Eq.\ (\ref{eq:zero-energy-spinor}) is an eigenstate with the zero eigenvalue of the Hamiltonian 
(\ref{eq:ham-A}), then it is also an eigenstate with the same eigenvalue of the square of this Hamiltonian. Using Eq.\ (\ref{eq:ham-A}), one then readily obtains 
\begin{equation}
H_{\rm eff}({\bf k},{\bf A})^2=B^2 ({\tilde{\bf k}}^2)^2+(1-2MB){\tilde{\bf k}}^2+M^2,
\end{equation}
with ${\tilde{\bf k}}\equiv{\bf k}+{\bf A}$, and the operator ${\tilde{\bf k}}^2$ after acting on the angular part of the upper component of the spinor (\ref{eq:zero-energy-spinor}) yields Eq.\ (\ref{eq:zero-mode-u}). Similarly, it may be shown that the function $v_l(r)$ in the spinor given by Eq.\ (\ref{eq:zero-energy-spinor}) obeys an equation of the same form as (\ref{eq:zero-mode-u}) with $l\rightarrow l+1$. From Eq.\ (\ref{eq:zero-mode-u}) we conclude that the function $u_{l-1}(r)$ is an eigenfunction of the operator ${\cal O}_{l-1/2}$ with a {\it positive} eigenvalue
\begin{equation}\label{eq:u}
{\cal O}_{l-\frac{1}{2}}u_{l-1}(r)=\lambda^2 u_{l-1}(r),
\end{equation}
since the operator ${\tilde{\bf k}}^2$ when acting on a function with the angular momentum $l$ is equal to $-{\cal O}_{l+1/2}$, and the eigenstates of the  operator ${\tilde{\bf k}}^2$
with a {\it negative} eigenvalue are localized.  Eqs.\ (\ref{eq:zero-mode-u}) and (\ref{eq:u}) then imply
\begin{equation}\label{eq:lambda-square}
\lambda_\pm=\frac{1\pm\sqrt{1-4MB}}{2B},
\end{equation}
and the function $u_l(r)\sim I_{l-\frac{1}{2}}(\lambda r)$ with $I_l(x)$ as the modified Bessel function of the first kind. However, from the above solutions the only square-integrable ones are in the zero angular-momentum channel since $I_l(x)\sim x^{-|l|}$ as $x\rightarrow0$. Furthermore, for $l=0$ only the linear combination $I_{1/2}(x)-I_{-1/2}(x)\sim x^{-1/2}e^{-x}$ has the asymptotic behavior at infinity consistent with a finite norm of the state.
In the above equation we should distinguish two regimes of parameters, $0<MB<1/4$ and $MB>1/4$, for which the argument of the square-root is positive and negative, respectively.

For $0<MB<1/4$, since the argument of the square-root in the above equation is positive, we obtain two zero-energy solutions
\begin{equation}
  \label{eq:zero-energy-states1}
\Psi_\pm({\bf r})=\frac{e^{-\lambda_\pm r}}{\sqrt{2\pi\lambda_\pm^{-1} r}}\left(\begin{array}{cc}e^{-i\varphi}\\ i\end{array}\right),
\end{equation}
and, of course, $\lambda_\pm>0$ because of the square-integrability. On the other hand, when $MB>1/4$, up to a normalization constant, the solutions have the form
\begin{eqnarray}\label{eq:zero-energy-states2}
\Psi_1({\bf r})&=&\frac{e^{-r\sqrt{\frac{M}{B}}\cos\theta}\cos\left(r\sqrt{\frac{M}{B}}\cos\theta\right)}{\sqrt{r}}\left(\begin{array}{cc}e^{-i\varphi}\\ i\end{array}\right),\nonumber\\
\Psi_2({\bf r})&=&\frac{e^{-r\sqrt{\frac{M}{B}}\cos\theta}\sin\left(r\sqrt{\frac{M}{B}}\cos\theta\right)}{\sqrt{r}}\left(\begin{array}{cc}e^{-i\varphi}\\ i\end{array}\right),
\end{eqnarray}
where
\begin{equation}
\theta=\frac{1}{2}\arctan\frac{\sqrt{|1-4MB|}}{1-2MB}.
\end{equation}
However, since the identity
\begin{equation}
  \label{eq:13}
  \sqrt{x}\cos{\left(\frac{1}{2}\arctan\frac{\sqrt{|1-4x|}}{1-2x}\right)}=\frac{1}{2}
\end{equation}
holds for $1/4<x<4$, the localization length of the zero-modes for $MB>1/4$ is actually independent of $M$, namely Eqs.~\eqref{eq:zero-energy-states2} become
\begin{eqnarray}
  \label{eq:zero-energy-states2final}
\Psi_1({\bf r})&=&\frac{e^{-\frac{r}{2B}}\cos\left(\frac{r}{2B}\right)}{\sqrt{r}}\left(\begin{array}{cc}e^{-i\varphi}\\ i\end{array}\right),\nonumber\\
\Psi_2({\bf r})&=&\frac{e^{-\frac{r}{2B}}\sin\left(\frac{r}{2B}\right)}{\sqrt{r}}\left(\begin{array}{cc}e^{-i\varphi}\\ i\end{array}\right).
\end{eqnarray}
Therefore, we can conclude that the Hamiltonian (\ref{eq:ham-A}) possesses zero-energy modes in the entire range of parameters $M$ and $B$ for which the system is in the topologically non-trivially phase, $0<M/B<4$. In particular, as it can be seen from Eq.\ (\ref{eq:lambda-square}), when $4MB<1$ zero-energy states are purely exponentially localized, while for $4MB>1$ the exponentially localized solutions have an oscillatory part with a characteristic length-scale exactly equal to the localization length.

Notice also that in the regime when $0<MB<1/4$, there are two characteristic length scales associated with the midgap modes, $\xi_\pm\sim\lambda_\pm^{-1}$. Of course, after a short-distance regularization is imposed, only a linear combination of the two states survives. The physical interpretation of the two length scales depends on the form of the superposition of the state after the regularization has been imposed, as it may be easily seen from the form of the states \eqref{eq:zero-energy-states1}. In the regime $MB>1/4$, the zero-energy states are characterized by a single length-scale $\xi_{\rm loc}\sim 2B$, which is at the same time the localization length and characterizes the oscillations of the exponentially decaying  state.

Therefore, the appearance of the zero-energy states bound to a $\pi$-flux vortex is a generic feature of the Hamiltonian (\ref{eq:cont-ham}) describing the quantum spin Hall system.  Furthermore, in the vortex-free system, it may be shown by imposing open boundary conditions on the wave-function at one of the edges of  the system, for instance the one perpendicular to the $x$-axis, and at infinity, $\Psi(x,y=0)=\Psi(x,y=\infty)=0$, that the Hamiltonian  gives rise to gapless edge modes with the penetration depth given by exactly the same expression as the localization length for the zero-energy modes bound to the $\pi$-flux vortex. The bulk-boundary correspondence may be thus probed by inserting a $\pi$-flux vortex in the quantum spin Hall system.

\section{Thin solenoid regularization of vortex}

The zero-energy modes, given by Eqs.\ (\ref{eq:zero-energy-states1}) and (\ref{eq:zero-energy-states2final}),
form an overcomplete basis in the zero angular-momentum channel, because the Hamiltonian (\ref{eq:ham-A}) is not
self-adjoint, which is due to the singularity of the vortex vector potential (\ref{eq:A}) at the origin. Thus the gauge potential has to be regularized.

A possible regularization is provided by considering the vortex with the flux concentrated in a thin annulus of a radius $R$. Let us first consider the Hamiltonian in the range of parameters $0<MB<1/4$. The zero-energy state of the Hamiltonian outside the annulus is then a linear combination of the modes $\Psi_\pm$ given by Eq.\ (\ref{eq:zero-energy-states1}). Inside the annulus the vector potential ${\bf A}=0$, and the zero-energy modes are
\begin{equation}\label{eq:vortex-free}
\Psi_{<}({\bf r})=C_1\left(\begin{array}{cc}e^{-i\varphi}I_1(\lambda_+r)\\ iI_0(\lambda_+r)\end{array}\right)+C_2\left(\begin{array}{cc}e^{-i\varphi}I_1(\lambda_{-}r)\\ iI_0(\lambda_-r)\end{array}\right),
\end{equation}
with $\lambda_\pm$ given by Eq.\ (\ref{eq:lambda-square}), and $C_{1,2}$ being complex constants. By matching these solutions at $r=R$, and taking $R\rightarrow0$, we obtain, up to a normalization constant, the zero-energy state of the form
\begin{equation}\label{eq:zero-energy-annulus}
\Psi({\bf r})=\frac{e^{-\lambda_+r}-e^{-\lambda_-r}}{\sqrt{r}}\left(\begin{array}{cc}e^{-i\varphi}\\ i\end{array}\right).
\end{equation}
Notice that this zero-energy state is regular at the origin which is a consequence of the regularity at the origin of the solutions (\ref{eq:vortex-free}) of the vortex-free problem. Similarly, one may show that when $MB>1/4$ the zero-energy mode is given by the spinor $\Psi_2$ in Eq.\ (\ref{eq:zero-energy-states2}) also regular at the origin and behaving $\sim r^{1/2}$ when $r\rightarrow0$.


\section{Self-adjoint extension of the Hamiltonian}
\label{sae}

Although the above regularization results in concrete solutions to the problem,
we should consider the self-adjoint extension of the corresponding Hamiltonian \eqref{eq:ham-A} in a more general manner by specifying the proper Hilbert space. That way, a family of Hermitian Hamiltonians is obtained, depending on free physical parameters that determine the scattering at the vortex core and the detailed profile of the single zero-mode (per spin). In the last section, we will comment on the regularization provided by the tight-binding version of the model, Eq.~\eqref{eq:upper-ham}.

The application of the standard theory of self-adjoint extensions (SAE) \cite{Weidmann,Thaller,Fulop:2007p3442,Jackiw:1991p1161} prescribes that we need to ensure that the massive Dirac Hamiltonian (a differential operator) becomes Hermitian (self-adjoint) only after choosing the proper Hilbert space (i.e. domain of functions) on which it is allowed to act. Instead of analyzing the imaginary spectrum (which needs to be removed), we implement von Neumann's construction by looking directly at the conditions under which the Hamiltonian is Hermitian when acting on arbitrary functions that are square integrable (but might diverge at the origin, due to diverging potential there). This will effectively determine the coefficients of the linear combination $C_1\Psi_++C_2\psi_-$ in angular momentum channel $l=0$ and thereby fix the zero-mode.

Using symmetry, we start from the radial part of the operator, $H^l(r)$, which acts in the subspace of angular momentum $l$ spanned by functions of the form
\begin{equation}
  \label{eq:4}
  \psi_l(r)\equiv e^{i l\varphi}\vect{e^{-i\varphi} u_l}{v_l},
\end{equation}
completely determined by $\svect{u_l}{v_l}$. Recall that the zero-energy states of the $M-B$ model \eqref{eq:ham-A} in presence of a $\pi$-flux vortex come in the form of Kramers pairs
\begin{equation}
  {\Psi}_{\uparrow}(x,y)=
\begin{pmatrix}
{\psi}(r,\varphi)\\
{0}
\end{pmatrix},
\qquad
  {\Psi}_{\downarrow}(r,\varphi)=
\begin{pmatrix}
{0}\\
{\psi}(r,\varphi)^*
\end{pmatrix}
\end{equation}
where ${\psi}(r,\varphi)$ is exactly of the form in Eq.~\eqref{eq:4}.

We also implement the standard change of scalar product in $r$-space by rescaling $\psi_l(r)=\frac{1}{\sqrt{r}}\tilde{\psi}_l(r)$, $\partial\psi_l(r)=\frac{1}{\sqrt{r}}\tilde{\partial}\tilde{\psi}_l(r)$ ($\partial$ always denotes $d/dr$), where $\tilde{\partial}\equiv\partial-\frac{1}{2r}$, after which $H^l$ takes the form
\begin{equation}
  \label{eq:5} \tilde{H}^l(r)=\mat{M+B\left(\partial^2-\frac{l(l-1)}{r^2}\right)}{-i\left(\partial+\frac{l}{r}\right)}{-i\left(\partial-\frac{l}{r}\right)}{-M-B\left(\partial^2-\frac{l(l+1)}{r^2}\right)}.
\end{equation}
We have reverted to the standard derivative ($\partial$) here. Since the $\pi$-flux enters through $l\rightarrow l_{eff}$, we omit it here. The non-derivative terms will not play any role in the following analysis since
the standard 'centrifugal force' provided by $l\neq 0$ does not lead to singularities. However, the gauge potential will have the chance to provide us with the boundary condition exactly when $l=0$.


Now, for two arbitrary wavefunctions $\phi$, $\psi$ which are determined by $\tilde{F}\equiv\svect{f}{g}$, $\tilde{U}\equiv\svect{u}{v}$ (we dropped index $l$), respectively, the condition of hermiticity of $\tilde{H}^l(r)$ becomes (note the change in $r\textrm{d}r$):
\begin{align}
  \label{eq:6}
  \bra{\tilde{\phi}}\tilde{H}-\tilde{H}^\dagger\ket{\tilde{\psi}}&= \int\!\textrm{d}r\; \tilde{\phi}(r)^*\tilde{H}^l(r)\tilde{\psi}(r)-  \left(\int\!\textrm{d}r\; \tilde{\psi}(r)^*\tilde{H}^l(r) \tilde{\phi}(r)\right)^*=\notag\\
  &=B\left.\left\{\tilde{f}^*\dr \tilde{u}-\dr \tilde{f}^*\tilde{u}-\tilde{g}^*\dr \tilde{v}+\dr \tilde{g}^*\tilde{v}\right\}\right|_0^\infty-\left.i\left\{\tilde{f}^*\tilde{v}+\tilde{g}^*\tilde{u}\right\}\right|_0^\infty=\notag\\
  &=B\left[\tilde{F}^*\sigma_3\dr \tilde{U}-\dr \tilde{F}^*\sigma_3 \tilde{U}-i \tilde{F}^*\sigma_1 \tilde{U}\right](0)\notag\\
  &\equiv 0,
\end{align}
where the $\sigma$ Pauli matrices act on the two component functions, which vanish at infinity, and are evaluated at the origin (point $r=0$) in the next-to-last line.

There is a continuous family of restrictions on the behavior of square-integrable functions at the origin, such that \eqref{eq:6} is satisfied, leading to the Hamiltonian which is Hermitian on such a chosen domain.
The proper parametrization of the most general restriction on the allowed domains is achieved by using the linearity of \eqref{eq:6}. Namely, we define two linear operators $\Gamma_1$, $\Gamma_2$ which map arbitrary functions, i.e. the domain of $\tilde{H}^\dagger$, onto their value at the boundary, i.e. the space of complex two component vectors:
\begin{equation}
  \label{eq:15}
\Gamma_i:\tilde{\psi}(r)\rightarrow\svect{\tilde{u}(0)}{\tilde{v}(0)}.
\end{equation}
These operators are defined by \eqref{eq:6}:
\begin{equation}
  \label{eq:7}
  B\left[\tilde{F}^*\sigma_3\dr \tilde{U}-\dr \tilde{F}^*\sigma_3 \tilde{U}-i \tilde{F}^*\sigma_1 \tilde{U}\right](0)\equiv \langle\Gamma_2\tilde{F},\Gamma_1\tilde{U}\rangle-\langle\Gamma_1\tilde{F},\Gamma_2\tilde{U}\rangle.
\end{equation}
Notice that this form can always be achieved due to the original form of the subtraction between $\tilde{H}$ and $\tilde{H}^\dagger$. We can choose in particular, without loss of generality,
\begin{align}
  \label{eq:8}
  \Gamma_1 \tilde{U}&=B\sigma_3\dr \tilde{U}(0)-i\frac{\sigma_1}{2}\tilde{U}(0),\\
  \Gamma_2 \tilde{U}&=\tilde{U}(0).
\end{align}
Any vector in the boundary space, i.e. $\svect{\chi_1}{\chi_2}\in\mathbb{C}^2$, is an image by $\Gamma_i$ of some wavefunction, i.e. of some $\svect{\tilde{u}}{\tilde{v}}$. Since $\tilde{U}(0)=\svect{\tilde{u}(0)}{\tilde{v}(0)}$ and $\dr\tilde{U}(0)=\svect{\dr\tilde{u}(0)}{\dr\tilde{v}(0)}$ take on arbitrary values, this means that the boundary space indeed is $\mathcal{H}_b=\mathbb{C}^2$.

The most general relation that has to be satisfied by a wavefunction such that \eqref{eq:6} will hold is now parametrized by unitary mappings $S$ in $\mathcal{H}_b$:
\begin{equation}
  \label{eq:9}
  \mathcal{D}(\tilde{H}_U)=\{\psi|(S-\sigma_0)\Gamma_1\psi+i(S+\sigma_0)\Gamma_2\psi=0\},
\end{equation}
with $\sigma_0$ the 2x2 identity matrix, and $\mathcal{D}$ denoting the domain of operator. One can directly understand from \eqref{eq:7} that forcing arbitrary linear combinations of a general $\tilde{U}(0)$ and $\dr\tilde{U}(0)$ to zero will still preserve the condition \eqref{eq:6}, due to the linearity and the antisymmetric nature of the form of this expression. Eq.~\eqref{eq:9} is giving us a precise recipe and parametrization of the fact that this is the most general restriction that needs to be made on the wavefunctions $\tilde{U}(r)$, i.e. on the $\psi(r)$.

We now proceed to use the form of $\Gamma_i$ to explore the allowed boundary conditions on the wavefunctions, in particular determining whether there is a self-adjoint extension $\tilde{H}_U$ with the previously found zero energy states in its domain. For concreteness we focus on the case $MB<1/4$.

Since $\mathcal{H}_b$ is $\mathbb{C}^2$, our mappings $S\in U(2)$, in contrast to the same problem in the case of an ordinary massive Dirac Hamiltonian necessitating a $U(1)$ parametrization \cite{2dDiracSAE}. The mappings $S\in U(2)$ can be parametrized by
\begin{equation}
  \label{eq:1}
  S=\frac{1}{d}\sum_\mu m_\mu\sigma_\mu,\quad m_\mu\in\mathbb{R},\quad\sum_\mu m_\mu^2=1,\quad d\equiv \matl{e^{i\eta}}{0}{0}{1},\quad \eta\in[0,2\pi),
\end{equation}
with the quaternion basis $\sigma_\mu=(\sigma_0,i \vec{\sigma})$. We will use the notation $[m_0,m_1,m_2,m_3]$ to represent the quaternion $\sum_\mu m_\mu\sigma_\mu$, while to label the boundary states of the zero energy wavefunctions $\psi_0\equiv C_1\Psi_++C_2\psi_-$ we will use
\begin{align}
  \label{eq:16}
\phi &\equiv\tilde{\psi_0}(0)\equiv\left.\sqrt{r}\psi_0\right|_{r=0}=\vect{C_1+C_2}{i(C_1+C_2)},\\ \phi' &\equiv\dr\tilde{\psi_0}(0)\equiv\left.\dr(\sqrt{r}\psi_0)\right|_{r=0}=-\vect{C_1\lambda_1+C_2\lambda_2}{i(C_1\lambda_1+C_2\lambda_2)}.
\end{align}
Due to normalization, both $\phi$ and $\phi'$ depend only on the vector $\svect{1}{x}$, where
\begin{equation}
  \label{eq:17}
  x\equiv\frac{C_1}{C_2}.
\end{equation}
It turns out that the case with $\eta=0$ is special, and so we examine it first in detail. Only this subclass contains the extension with regular zero-modes.

\subsection{Extensions described by $SU(2)$, $\eta=0$}

Eqs.~\eqref{eq:9},~\eqref{eq:1} lead to the following condition on two quaternions $P,Q$:
\begin{align}
  \label{eq:2}
  B\cdot P\phi'&=Q\phi,\\
  Q&=[\frac{m_1}{2}+i(1+m_0),\frac{1-m_0}{2}+i m_1,\frac{m_3}{2}+i m_2,-\frac{m_2}{2}+i m_3], \notag\\
  P&=[-i m_3, -i m_2, i m_1, i(m_0-1)].
\end{align}
Once the values of $M,B$ (and therefore also $\lambda_{1/2}\equiv\lambda_\pm$) are given, this equation determines $x$ as function of the particular SAE $m_\mu$ (if a solution for $x$ exists), and $x$ then determines the specific linear combination of $\lambda_{1/2}$ decaying functions in the zero mode, influencing also its regularity properties at the origin. Since $\mathrm{det}(P)=2(m_0-1)$, we first consider:

\subsubsection{Extension with $\eta=0$, $m_0=1$}
\label{saeregular}

According to definition \eqref{eq:1} this immediately implies $\vec{m}=0$, and
\begin{align}
  \label{eq:18}
  \sigma_0\phi=0&\Longleftrightarrow \sqrt{r}\psi_0|_{r=0}=0,\text{ i.e.}\\
  C_1&=-C_2.
\end{align}
Such a wavefunction is regular at the origin and localized on the scale $1/\textrm{min}(\lambda_i)$, no matter the values $M,B$ (or $\lambda_{1,2}$). We therefore see that such a wavefunction is allowed when the physical SAE is given by $S=[1,0,0,0]$, $\eta=0$. This will turn out to be the only extension allowing $x=-1$, see Fig.~\ref{fig:1}, essentially because it is the only extension for which the matrix on the left-hand side of Eq.~(\ref{eq:2}) vanishes.

\subsubsection{Extension with $\eta=0$, $m_0<1$}

We must now consider Eq.~\eqref{eq:2} as a vector equation, treating $C_{1,2}$ as unknown variables. We get
\begin{align}
  \label{eq:3}
  \hat{Y}\vect{C_1}{C_2}&=0\\
  \hat{Y}&=\matl{Ba\lambda_1+p}{Ba\lambda_2+p}{Bb\lambda_1+q}{Bb\lambda_2+q}\\
  \textrm{det}(\hat{Y})&=B(\lambda_1-\lambda_2)(aq-bp),
\end{align}
where $a\equiv P_{11}+i P_{12},\ b\equiv P_{21}+i P_{22},\ p\equiv Q_{11}+i Q_{12},\ q\equiv Q_{21}+i Q_{22}$. There are solutions only when
\begin{equation}
  \label{eq:19}
  aq=bp\Leftrightarrow m_2=0.
\end{equation}
So in the case of $m_0\neq 1$, the zero energy mode is allowed in principle only for Hamiltonians with $m_2=0$. The value of $x$ in such a system is given by
\begin{equation}
  \label{eq:10}
  x=-\frac{g(\lambda_2)}{g(\lambda_1)}, \quad g(\lambda)\equiv B a\lambda+p=B b\lambda+q,
\end{equation}
where
\begin{equation}
  \label{eq:11}
  g(\lambda)=B(1-m_0-m_1-i m_3)\lambda+\left(\frac{1}{2}-i\right)(m_1-1-i m_3)+\left(\frac{1}{2}+i\right)m_0
\end{equation}
is a linear function of $\lambda$. From this expression it seems that another way for $x=-1$ (and the zero mode becoming regular at the origin for all model parameters $M,B$), is if $m_1=1$, where we assume $\lambda_1\neq\lambda_2$. However, this is not the case, because when $m_1=1$ only one component of the vector equation Eq.~(\ref{eq:3}) can be satisfied, and not both component equations at the same time.

Note that the extensions identified here by having $m_2=0$ include the case from previous subsection, $m_0=1$. In fact, demanding that $x=-1$ directly from Eq.~(\ref{eq:3}) implies that the columns of $Y$ must be the same, which reduces to the demand $a=b=0\Rightarrow P=[0,0,0,0]$ since $\lambda_{1,2}$ are non-degenerate. It is easy to see, using the definition Eq.~(\ref{eq:1}), that this is only possible in the above considered case $m_0=1$.

\subsection{Dependence on $MB$}

\begin{figure}
  \centering
\includegraphics[width=1\textwidth]{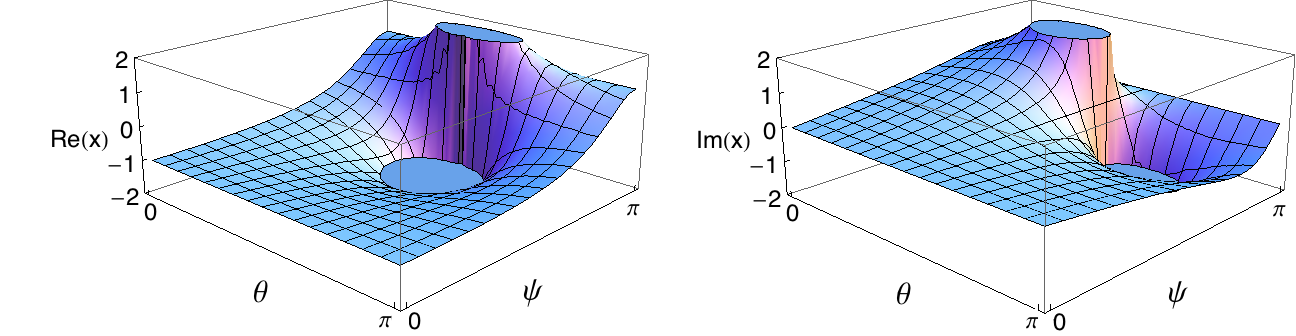}
\caption{ \label{fig:1} Vortex singularity regularization. The value of $x=C_1/C_2$, where the zero-mode wavefunction $\Psi_0\sim r^{-1/2}(x e^{-\lambda_1 r}+e^{-\lambda_2 r})$, is presented as a function of the SAE parametrized by two angles $(\psi,\theta)$, and $\eta=0$ (see Section~\ref{sae}). The wavefunction becomes regular at the origin for $x=-1$. Notice that Re$(x)$ approaches $-1$ for $\psi=0$, where also Im$(x)=0$. The plots are evaluated for $MB=0.15$, but are representative for all topologically non-trivial model parameters $0<MB<1/4$.}
\end{figure}

To analyze possible values for $x$ for given model parameters $M,B$, we parametrize the SAE using hyperspherical coordinates
\begin{equation}
  \label{eq:14}
m_\mu=(\cos{\psi},\sin{\psi}\cos{\theta},\sin{\psi}\sin{\theta}\cos{\varphi},\sin{\psi}\sin{\theta}\sin{\varphi}),
\end{equation}
where the hypersphere $SU(2)\simeq S^3$ is properly covered by $\psi\in[0,\pi]$, $\theta\in[0,\pi]$, and $\varphi\in[0,2\pi)$. The condition $m_2=0$, Eq.~(\ref{eq:19}), means $\varphi=\pi/2$, and also $\psi\neq 0$ ($m_0\neq 1$), $(\psi,\theta)\neq(\frac{\pi}{2},0)$ ($m_1\neq 1$) in this subsection. This parametrization can be plugged into the function $g(\lambda)$ from Eq.~\eqref{eq:11}
together with the values of $\lambda_{1/2}=\frac{1\pm\sqrt{1-4k}}{2}$, where $k\equiv MB$, so that one can deduce $x$ from the model parameters $M,B$ and the boundary condition determined by $\psi,\theta$. Analysis shows that $x(\psi,\theta)$ weakly depends on the value of $k\equiv MB$, and so Fig.~\ref{fig:1} presents the typical behavior for fixed $k$. On the $(\psi,\theta)$ plane, the function $x(\psi,\theta)$ is strongly localized and a sharply peaked dipole. One can use the real and imaginary parts of $x$ to identify structure in the scattering phase shift coming from the regularized vortex.

\subsection{From $SU(2)$ to $U(2)$ extensions, $\eta\neq 0$}

The introduction of $\eta\neq 0$ in Eq.~(\ref{eq:1}) does not change the analysis method of previous subsections.

To start, the expressions for quaternions $P,Q$ in Eq.~(\ref{eq:2}) are slightly more complicated, so that demanding $x=-1$ through the vanishing of left-hand side of Eq.~(\ref{eq:2}), as in Eq.~(\ref{eq:18}), becomes
\begin{align}
  \label{eq:20}
  \mathrm{det}(P)&\equiv 0,\text{ i.e.}\\\notag
  \exp(i\eta)&=\frac{m_3+i(1-m_0)}{m_3-i(1-m_0)}.
\end{align}
However, even when the determinant vanishes, the matrix $P$ can not become zero matrix, unless $\cos(\eta)=1\Rightarrow\eta=0$, which actually just takes us back to a SAE considered in previous subsections.

Next, considering Eqs.~(\ref{eq:2}) by treating $C_1,C_2$ as unknowns, as in Eq.~(\ref{eq:3}), implies that zero-modes can exist only for a SAE that has
\begin{align}
  \label{eq:21}
  \mathrm{det}(Y)&=0\Longrightarrow\\
  \exp(i\eta)&=\frac{1+m_1-im_2}{1+m_1+im_2}\Longleftrightarrow\\
  \tan\left(\frac{\eta+n\pi}{2}\right)&=\frac{\sin(\psi) \sin(\theta) \cos(\phi)}{1+\sin(\psi) \cos(\theta)},
\end{align}
where in the last line we used the parametrization from Eq.~(\ref{eq:14}).

We emphasize that here, as for $\eta=0$ case, demanding a zero-mode with $x=-1$ implies from Eq.~(\ref{eq:3}) that the columns of $Y$ have to be the same. Since $\lambda_{1,2}$ are not degenerate for physical values of $MB$, this gives $a=b=0$, i.e. the already considered demand that the matrix $P$ vanishes, which is impossible when $\eta\neq 0$.

Following the derivation of Eq.~(\ref{eq:10}), the $x$ in general is again given by a ratio
\begin{equation}
  \label{eq:22}
x=-\frac{\tilde{g}(m_\mu,\eta,\lambda_2)}{\tilde{g}(m_\mu,\eta,\lambda_1)},
\end{equation}
where we omit the lengthy explicit expression for $\tilde{g}$. Notice that even when this ratio approaches a singular limit by $\tilde{g}\rightarrow 0\text{ or }\infty$, the previous paragraph ensures that $x$ will not reach the value $-1$.

To summarize, the SAE analysis of the $M-B$ model shows that a $U(2)$ parametrization which completes this model allows the existence of zero modesfor a subclass of Hamiltonians described by three free angle parameters, e.g. $\eta,\theta,\psi$ in Eq.~\eqref{eq:14}. These determine a matrix relating the value of the wavefunction spinor to its derivative, both taken at the origin, according to Eq.~\eqref{eq:2}. There is only a single SAE, Section \ref{saeregular}, determined by $\eta=\psi=0$, which allows for the existence of a zero energy state which is \textit{both} localized and regular (vanishing) at the origin.

\section{Comparison to the tight-binding $M-B$ model}
\label{sec:tb}

\begin{figure}[!h]
  \centering
\includegraphics[width=1\textwidth]{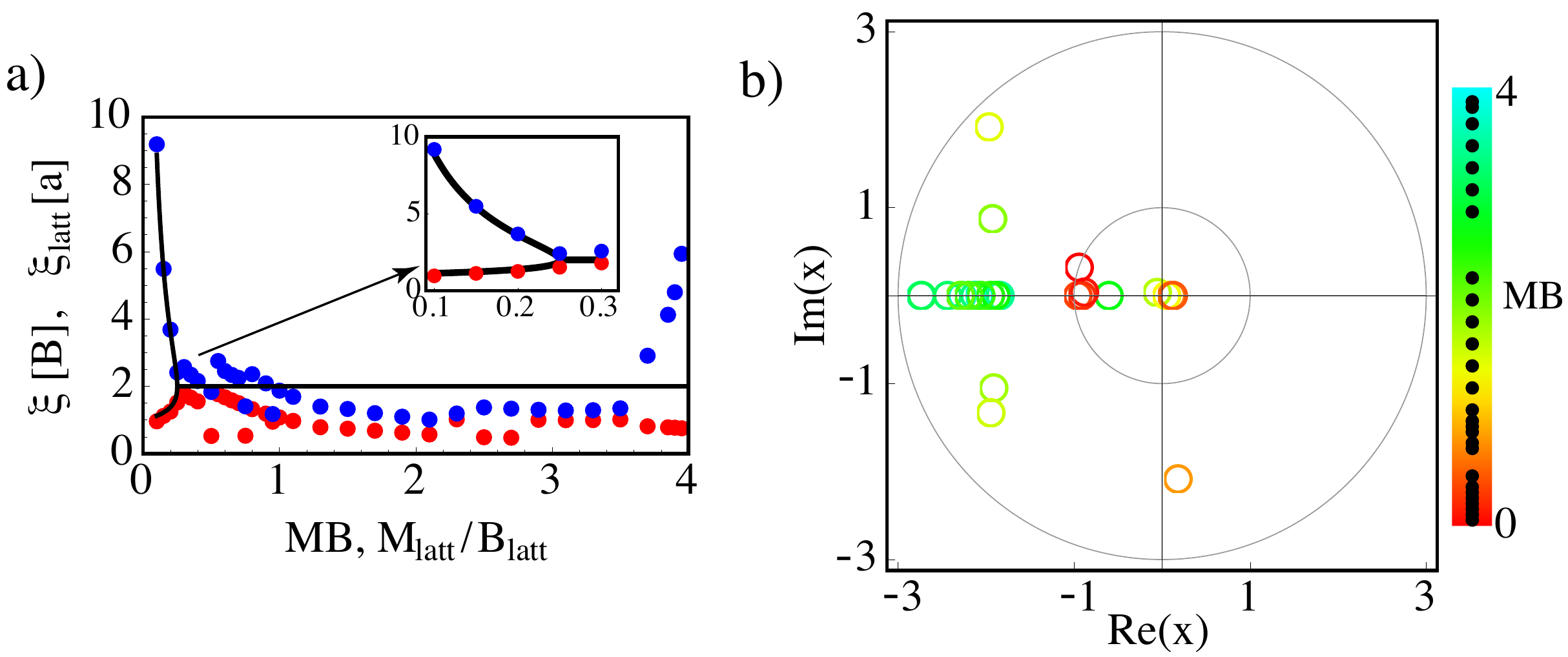}
\caption{ \label{fig:2} a) Comparison of the vortex-bound zero-mode localization lengths $\xi_\pm$ (black lines) predicted by the present $M-B$ model (see text after Eq.~\eqref{eq:zero-energy-states2final}), and measured in the tight-binding version of the model on a 31x31 lattice. Inset shows in detail the excellent agreement in the $MB<1/4$ regime.
The lattice model has a phase transition at $M_{latt}/B_{latt}=4$ which is absent in the continuum model, since the gap closes at finite momentum. b) The complex constant $x=C_1/C_2$ determines the form of the zero-mode wavefunction. When $x=-1$ the mode is regular at the $\pi$-flux, and this situation is realized throughout the regime $MB<1/4$ (dots in the legend mark the values of $MB$ of the plotted points).}
\end{figure}

In this Section we present the results from a tight-binding $M-B$ model, which in momentum space has the limit Eqs.~\eqref{eq:tight-binding},~\eqref{eq:upper-ham}. We numerically study this model on a 31x31 sized square lattice with periodic boundary conditions and a $\pi$-flux---anti-$\pi$-flux pair positioned at maximal distance. The details of this tight-binding model written in real space, i.e. on the lattice, are presented in Ref.~\cite{TopProbes}.

We first isolate the zero energy mode localized on a single $\pi$-flux, using the nearly degenerate inversion-symmetric and anti-symmetric zero modes on the finite lattice. The lattice symmetries guarantee the $(\exp{(-i\phi)},i)^T$ form of the wavefunction spinor, with $\phi$ the polar angle, just as in the continuum. By fitting the radial envelope of the spinor using Eq.~(\ref{eq:zero-energy-states1}) and the angle-averaged wavefunction, we extract the two localization lengths $\xi^\pm_{latt}$ in units of the lattice constant. The results are shown in Fig.~\ref{fig:2}(a). The agreement with the present continuum model prediction, see text after Eq.~\eqref{eq:zero-energy-states2} and Eq.~\eqref{eq:lambda-square}, is excellent in the $MB<1/4$ regime (inset of Fig.~\ref{fig:2}(a)). For $MB>1/4$ the oscillatory part of the wavefunction (see Eq.~\eqref{eq:zero-energy-states2final}) makes the numerical fitting less reliable, and the agreement is only qualitative.

One should note that the lattice tight-binding model has a natural length-scale, the lattice constant $a$, beside its dimensionless constants $M_{latt},B_{latt}$ (in units of inter-site hopping energy $A$), see Eq.~\eqref{eq:tight-binding}. In the present continuum model, Eq.~\eqref{eq:ham-A}, the parameters $M,B$ determine two length-scales, $B$ and $B^{1/2}M^{-1/2}$, where the former acts as a length-scale in the zero-modes. The lattice constant is simply absorbed by $M_{latt},B_{latt}$ to give $M,B$ of the continuum model (see Eq.~\eqref{eq:cont-ham}), and that is why the agreement in Fig.~\ref{fig:2}(a) is quantitatively precise. (In the lattice calculation we actually set $B_{latt}\equiv 1$ and vary only $M_{latt}$, driving therefore both the parameter $MB=M_{latt}B_{latt}$ and the lattice topological transition parameter $M_{latt}/B_{latt}$.)

To determine the SAE which is realized in the lattice model, we calculate the complex constant $x=C_1/C_2$, which was defined in Eq.~(\ref{eq:17}), and represents the ratio of contributions of two singular functions in the zero-mode wavefunction, Eqs.~(\ref{eq:zero-energy-states1}),~(\ref{eq:zero-energy-states2final}). Fig.~\ref{fig:2}(b) shows that throughout the regime $MB<1/4$, $x$ keeps near the value $x=-1$, which is the special case of zero-mode being regular at the $\pi$-flux position. As shown above, only a single SAE allows this, and this SAE is also realized in a thin-solenoid regularization of the continuum model.


\section{Quantum numbers of the zero-energy modes}

We will now show that these zero-energy modes carry non-trivial charge or spin quantum number depending on their occupation. For that purpose, we will write the continuum $4\times 4$ Hamiltonian (\ref{eq:cont-ham}) coupled to a $U(1)$ vector potential as
\begin{equation}
H=i\gamma_0\gamma_i(k_i+A_i)+(M-B({\bf k}+{\bf A})^2)\gamma_0
\end{equation}
with the vector potential ${\bf A}$ given by Eq.\ (\ref{eq:A}). Note that the unitary matrices $\gamma_3=\sigma_2\otimes\tau_2$ and $\gamma_5=\sigma_2\otimes\tau_1$ anticommute with the gamma-matrices $\gamma_\alpha$, $\alpha=0,1,2$. Therefore, the Hamiltonian anticommutes with the matrices $\Gamma_3\equiv i\gamma_0\gamma_3$ and $\Gamma_5\equiv i\gamma_0\gamma_5$ which then generate chiral (spectral) symmetry relating states with positive and negative energies, i.e., if $H|E\rangle=E|E\rangle$, then, for instance, $\Gamma_5|E\rangle=|-E\rangle$, and the matrix $\Gamma_5$ reduces in the zero-energy subspace of the Hamiltonian. These two properties then imply that in the ground-state the expectation value of a traceless Hermitian operator $Q$ is given in terms of the zero-energy states of the Hamiltonian
\begin{equation}
\langle Q\rangle=\frac{1}{2}\left(\sum_{{\rm occupied}}\Psi^\dagger Q \Psi-\sum_{{\rm unoccupied}}\Psi^\dagger Q \Psi\right).
\end{equation}
Therefore, in the case of a quantum spin Hall insulator threaded by a $\pi$-flux vortex, depending on the occupation of a pair of zero-energy states, there are four possibilities for the ground state quantum numbers. Namely, when both states are occupied or empty, according to the above expression, the charge is $+e$ or $-e$ and the spin quantum number is zero. On the other hand, when one of the states is occupied, the spin quantum number is $+1/2$ or $-1/2$, while the charge is zero. In that way, the spin-charge separation, characteristic for one-dimensional systems\cite{SSH}, appears also in a two-dimensional system \cite{hou-chamon}, and is tied to a topologically non-trivial nature of the quantum spin Hall state \cite{dung-hai1,qi-zhang}. Similarly, zero-energy modes bound to the vortex core in an antiferromagnetic state on a honeycomb lattice lead to the phenomenon of spin fragmentation \cite{igor-spin-fragmentation}; see Ref.\ \cite{igor-Dirac-isospin} for a general discussion of this class of problems in the context of Dirac systems.

Finally, let us note that the topological stability of $\pi$-flux zero-modes, i.e. stability under smooth deformations of the vector potential, has been strictly proved for both Dirac and Schr\" odinger Hamiltonians \cite{aharonov-casher,jackiw-prd84}, but for the $M-B$ model, which is a sum of both, we are not aware of an analogous proof based on an index theorem.


\section{Conclusions}

In conclusion, we have shown that a pair of zero-modes bound to a magnetic $\pi$-flux are a generic feature of the $M-B$ model in the topologically non-trivial phase. The continuum Hamiltonian of that model has both Dirac and Schr\" odinger kinetic terms and a mass, and has not been studied in detail previously. We have analytically found the zero-modes in the presence of the $\pi$-flux vortex in the entire range of parameters describing a topologically non-trivial phase with the bandgap opening at the zero momentum. These modes are exponentially localized around the vortex core, and a particular regularization of the vector potential corresponding to the vortex yields the modes regular at the origin, but in general, as we have shown, the form of the solution depends on the short-distance regularization of the vortex. Vortex zero modes obtained within a lattice tight-binding model match the ones found for a particular self-adjoint extension of the continuum Hamiltonian. Finally, we discussed a realization of the two-dimensional spin-charge separation through the vortex zero-modes.

\section*{Acknowledgements}

The authors gladly thank Oskar Vafek and Igor Herbut for useful discussions. This work was supported by Dutch Foundation for Fundamental Research on Matter (FOM). V.\ J.\ acknowledges the support of the Netherlands Organization for Scientific Research (NWO).








\end{document}